\documentclass[letterpaper, 11pt]{article}
\usepackage{amssymb}
\usepackage{amsthm}
\usepackage{amsmath}
\usepackage{fullpage}
\usepackage{epsfig}
\usepackage{hyperref,txfonts,mathrsfs}

\DeclareSymbolFont{symbolsC}{U}{txsyc}{m}{n}
\SetSymbolFont{symbolsC}{bold}{U}{txsyc}{bx}{n}
\DeclareFontSubstitution{U}{txsyc}{m}{n}
\DeclareMathSymbol{\coloneqq}{\mathrel}{symbolsC}{66}

\newcommand\remove[1]{}

\newcommand{\rnote}[1]{}

\newcommand{\1}{\mathbf{1}}

\newcommand{\R}{\mathbb{R}}

\newcommand{\e}{\varepsilon}

\newcommand{\h}{\mathscr{H}}

\newcommand{\E}{\mathbb{E}}

\newtheorem{theorem}{Theorem}[section]
\newtheorem{lemma}[theorem]{Lemma}
\newtheorem{prop}[theorem]{Proposition}

\newtheorem{corollary}[theorem]{Corollary}

\newtheorem{definition}[theorem]{Definition}

\begin{document}


\title{Lower bounds on Locality Sensitive Hashing}
\author{Rajeev Motwani \and Assaf Naor\and Rina Panigrahy}


\date{}
\maketitle

\begin{abstract}
Given a metric space $(X,d_X)$, $c\ge 1$, $r>0$, and $p,q\in [0,1]$,
a distribution over mappings $\h:X\to \mathbb N$ is called a
$(r,cr,p,q)$-sensitive hash family if any two points in $X$ at
distance at most $r$ are mapped by $\h$ to the same value with
probability at least $p$, and any two points at distance greater
than $cr$ are mapped by $\h$ to the same value with probability at
most $q$. This notion was introduced by Indyk and Motwani in 1998 as
the basis for an efficient approximate nearest neighbor search
algorithm, and has since been used extensively for this purpose. The
performance of these algorithms is governed by the parameter
$\rho=\frac{\log(1/p)}{\log(1/q)}$, and constructing hash families
with small $\rho$ automatically yields improved nearest neighbor
algorithms. Here we show that for $X=\ell_1$ it is impossible to
achieve $\rho\le \frac{1}{2c}$. This almost matches the construction
of Indyk and Motwani which achieves $\rho\le \frac{1}{c}$.
\end{abstract}

\section{Introduction}

In this note we study the complexity of finding the nearest neighbor
of a query point in certain high dimensional spaces using {\em
Locality Sensitive Hashing} (LSH). The nearest neighbor problem is
formulated as follows: Given a database of $n$ points in a metric
space, preprocess it so that given a new query point it is possible
to quickly find the point closest to it in the data set. This
fundamental problem arises in numerous applications, including data
mining, information retrieval, and image search, where distinctive
features of the objects are represented as points in $\mathbb R^d$.
There is a vast amount of literature on this topic, and we shall not
attempt to discuss it here. We refer the interested reader to the
papers~\cite{InM98,Har01,DIIM04,Pan06}, and especially to the
references therein, for background on the nearest neighbor problem.

While the exact nearest neighbor problem seems to suffer from the
``curse of dimensionality'', many efficient techniques have been
devised for finding an approximate solution whose distance from the
query point is at most $c$ times its distance from the nearest
neighbor. One of the most versatile and efficient methods for
approximate nearest neighbor search is based on Locality Sensitive
Hashing, as introduced by Indyk and Motwani in 1998~\cite{InM98}.
This method has been refined and improved in several papers- the
most recent algorithm can be found in~\cite{DIIM04}. We also refer
the reader to the LSH website, where more information on this
algorithm can be found, including its implementation and code- all
this can be found at  \url{
http://web.mit.edu/andoni/www/LSH/index.html}. The LSH approach to
the approximate nearest neighbor problem is based on the following
concept.

\begin{definition}\label{def:lsh} Let $(X,d_X)$ be a metric space, $r,R>0$ and
$p,q\in [0,1]$. A distribution over mappings $\h:X\to \mathbb N$ is
called a $(r,R,p,q)$-sensitive hash family if for any $x,y\in X$,

\begin{itemize}
\item $d_X(x,y)\le r\implies\Pr[\h(x)=\h(y)]\ge p$\enspace.

\item $d_X(x,y)>R\implies\Pr[\h(x)=\h(y)]\le q$\enspace.
\end{itemize}
\end{definition}

Given $c\ge 1$ we define
\begin{eqnarray}\label{eq:def rho}
\rho_X(c)=\sup_{r>0}\inf\left\{\frac{\log(1/p)}{\log(1/q)}:\ \exists
(r,cr,p,q)-\mathrm{sensitive\ hash\ family}\ \h:X\to \mathbb
N\right\}\enspace.
\end{eqnarray}
Of particular interest is the case $X=\ell_s^d$, for some $s\ge 1$
and $d\in \mathbb N$. In this case we define
\begin{eqnarray*}\label{eq:def s}
\rho_s(c)=\limsup_{d\to\infty} \rho_{\ell_s^d}(c)\enspace.
\end{eqnarray*}

The importance of these parameters stems from the following
application to approximate nearest neighbor search. It will be
convenient to discuss it in the framework of the following decision
version of the $c$-approximate nearest neighbor problem: Given a
query point, find any element of the data set which is at distance
at most $cr$ from it, provided that there is a data point at
distance at most $r$ from the query point. This decision version is
known as the $(r, cr)$-near neighbor problem. It is well known that
the reduction to the decision version adds only a logarithmic factor
in the time and space complexity~\cite{InM98, Har01}. The following
theorem was proved in~\cite{InM98}; the exact formulation presented
here is taken from~\cite{DIIM04}.

\begin{theorem} Let $(X,d_X)$ be a metric on a subset of $\R^d$.
Suppose that $(X,d_X)$ admits a $(r,cr,p,q)$-sensitive hash family
$\h$, and write $\rho=\frac{\log(1/p)}{\log(1/q)}$. Then for any
$n\ge \frac{1}{q}$ there exists a randomized algorithm for $(r,c)$
near neighbor on $n$-point subsets of $X$ which uses
$O\left(dn+n^{1+\rho}\right)$ space, with query time dominated by
$O\left(n^\rho\right)$ distance computations and
$O\left(n^\rho\log_{1/q} n\right)$ evaluations of hash functions
from $\h$.
\end{theorem}

Thus, obtaining bounds on $\rho_X(c)$ is of great algorithmic
interest. It is proved in~\cite{InM98} that $\rho_1(c)\le 1/c$, and
for small values of $c$, namely $c\in [1,10]$, is was shown
in~\cite{DIIM04} that this inequality is strict. We refer
to~\cite{DIIM04} for numerical data on the best know estimates for
$\rho_1(c)$ for small $c$. For $s=2$ a recent result of Andoni and
Indyk~\cite{AI05} shows that $\rho_2(c)\le 1/c^2$, and for general
$s\in [1,2]$ the best known bounds~\cite{DIIM04} are $\rho_s(c)\le
\max\{1/c,1/c^s\}$.

The main purpose of this note is to obtain lower bounds on
$\rho_1(c)$ and $\rho_2(c)$ which nearly match the bounds obtained
from the constructions in~\cite{InM98,DIIM04,AI05}. Our main result
is:

\begin{theorem}\label{thm:main} For every $c,s\ge 1$,
\begin{eqnarray}\label{eq:main}
\rho_s(c)\ge\frac{e^{\frac{1}{c^s}}-1}{e^{\frac{1}{c^s}}+1}\ge
\frac{e-1}{e+1}\cdot \frac{1}{c^s}\ge \frac{0.462}{c^s}\enspace .
\end{eqnarray}
\end{theorem}

The second to last inequality in~\eqref{eq:main} follows from
concavity of the function $t\mapsto \frac{e^t-1}{e^t+1}$ on
$[0,\infty)$. Observe also that as $c\to \infty$,
$\frac{e^{1/c}-1}{e^{1/c}+1}\sim \frac{1}{2c}$. It would be very
interesting to determine $\limsup_{c\to\infty}c\cdot\rho_1(c)$
exactly- due to Theorem~\ref{thm:main} and the results
of~\cite{InM98} we currently know that this number is in the
interval $[1/2,1]$.

\section{Proof of Theorem~\ref{thm:main}}

The basic idea in the proof of Theorem~\ref{thm:main} is simple.
Choose a random point $x\in \{0,1\}^d$ and consider the random
subset $A$ of the cube $\{0,1\}^d$ consisting of points $u$ for
which $\h(u)=\h(x)$. The second condition in
Definition~\ref{def:lsh} forces $A$ to be small in expectation. But,
when $A$ is small we can bound from above the probability  that
after $r$ steps, the random walk starting at a random point in $A$
will end up in $A$. We obtain this upper bound using a Fourier
analytic argument, and in combination with the first condition in
Definition~\ref{def:lsh} we deduce the desired bound on $\rho_1(c)$.

Theorem~\ref{thm:main} follows from the following result:

\begin{prop}\label{prop:general} Let $\h$ be a $(r,R,p,q)$-sensitive hash family on the
Hamming cube $(\{0,1\}^d,\|\cdot\|_1)$. Assume that $r$ is an odd
integer and that $R<\frac{d}{2}$. Then
$$
p\le
\left(q+e^{-\frac{1}{d}\left(\frac{d}{2}-R\right)^2}\right)^{\frac{e^{2r/d}-1}{e^{2r/d}+1}}\enspace.
$$
\end{prop}

Choosing $R\approx \frac{d}{2}-\sqrt{d\log d}$ and $r\approx R/c$ in
Proposition~\ref{prop:general}, and letting $d\to \infty$, yields
Theorem~\ref{thm:main} in the case $s=1$. The case of general $s\ge
1$ follows from the fact that for $x,y\in \{0,1\}^d$,
$\|x-y\|_s=\|x-y\|_1^{1/s}$.

The proof of Proposition~\ref{prop:general} will be broken into a
few lemmas.

\begin{lemma}\label{lem:count} Let $\h$ be a $(r,R,p,q)$-sensitive hash family on the
Hamming cube $(\{0,1\}^d,\|\cdot\|_1)$, and fix $x\in \{0,1\}^d$.
Then
$$
\E \left|\h^{-1}\left(\h(x)\right)\right|\le \sum_{k=0}^{\lfloor
R\rfloor} \binom{d}{k}+ q \cdot \sum_{k=\lfloor R\rfloor+1}^{d}
\binom{d}{k}\enspace.
$$
\end{lemma}

\begin{proof} We simply write
\begin{eqnarray*}
\E \left|\h^{-1}\left(\h(x)\right)\right|&=&\sum_{u\in \{0,1\}^d}
\Pr[\h(u)=\h(x)]\\ &\le& \left|\{u\in \{0,1\}^d:\ \|u-x\|_1\le
R\}\right|+ q \cdot\left|\{u\in \{0,1\}^d:\
\|u-x\|_1> R\}\right|\\
&=& \sum_{k=0}^{\lfloor R\rfloor} \binom{d}{k}+ q \cdot
\sum_{k=\lfloor R\rfloor+1}^{d} \binom{d}{k}\enspace.
\end{eqnarray*}
\end{proof}

\begin{corollary}\label{coro:use} Assume that
$R<\frac{d}{2}$. Then, using the notation of Lemma~\ref{lem:count},
we have that
$$
\E\left|\h^{-1}\left(\h(x)\right)\right|\le
2^d\left(q+e^{-\frac{1}{d}\left(\frac{d}{2}-R\right)^2}\right)\enspace.
$$
\end{corollary}

\begin{proof} This follows from Lemma~\ref{lem:count} and the standard estimate
$ \sum_{k\le \frac{d}{2}-a} \binom{d}{k}\le 2^d\cdot
e^{-\frac{a^2}{d}}. $
\end{proof}

\begin{lemma}[Random walk lemma]\label{lem:walk} Let $r$ be an odd integer.
Given $\emptyset \neq B\subseteq \{0,1\}^d$, consider the random
variable $Q_B\in \{0,1\}^d$ defined as follows: Choose a point $z\in
B$ uniformly at random, and perform $r$-steps of the standard random
walk on the Hamming cube starting from $z$. The point thus obtained
will be denoted $Q_B$. Then
$$
\Pr[Q_B\in B]\le
\left(\frac{|B|}{2^d}\right)^{\frac{e^{2r/d}-1}{e^{2r/d}+1}}\enspace.
$$
\end{lemma}

\begin{proof} We begin by recalling some background and notation on
Fourier analysis on the Hamming cube. Given $S\subseteq \{1,\ldots
d\}$, the Walsh function $W_S:\{0,1\}^d\to \{-1,1\}$ is defined by
$$
W_S(u)=(-1)^{\sum_{j\in S} u_j}\enspace.
$$

For $f:\{0,1\}^d\to \R$ we set
$$
\widehat f (S)=\frac{1}{2^d} \sum_{u\in \{0,1\}^d}
f(u)W_S(u)\enspace,
$$
so that $f$ can be decomposed as follows:
$$
f=  \sum_{S\subseteq\{1,\ldots,d\}} \widehat f(S)W_S\enspace.
$$
For every $f,g:\{0,1\}^d\to \R$ we write
$$
\langle f,g\rangle= \frac{1}{2^d} \sum_{u\in \{0,1\}^d}
f(u)g(u)\enspace.
$$
By Parseval's identity,
$$
\langle f,g\rangle=  \sum_{S\subseteq\{1,\ldots,d\}} \widehat
f(S)\widehat g(S)\enspace.
$$

For $\e\in [0,1]$ the Bonami-Beckner operator $T_\e$ is defined as
$$
T_\e f= \sum_{S\subseteq\{1,\ldots,d\}} \e^{|S|}\widehat
f(S)W_S\enspace.
$$
The Bonami-Beckner inequality~\cite{Bonami70,Beck75} states that for
every $f:\{0,1\}^d\to \R$,
$$
\sum_{S\subseteq\{1,\ldots,d\}} \e^{2|S|}\widehat f(S)^2=\|T_\e
f\|_2^2=\frac{1}{2^d} \sum_{u\in \{0,1\}^d} \left(T_\e
f(u)\right)^2\le \|f\|_{1+\e^2}^2=\left(\frac{1}{2^d} \sum_{u\in
\{0,1\}^d}f(u)^{1+\e^2}\right)^{\frac{2}{1+\e^2}}\enspace.
$$
Specializing to the indicator of $B\subseteq \{0,1\}^d$ we get that
\begin{eqnarray}\label{eq:use}
\sum_{S\subseteq\{1,\ldots,d\}} \e^{2|S|}\widehat {\1_B}(S)^2\le
\left(\frac{|B|}{2^d}\right)^{\frac{2}{1+\e^2}}\enspace.
\end{eqnarray}

Now, let $P$ be the transition matrix of the standard random walk on
$\{0,1\}^d$, i.e. $P_{uv}=1/d$ if $u$ and $v$ differ in exactly one
coordinate, $P_{uv}=0$ otherwise. By a direct computation we have
that for every $S\subseteq \{1,\ldots, d\}$,
$$
PW_S=\left(1-\frac{2|S|}{d}\right)W_S\enspace,
$$
i.e. $W_S$ is an eigenvector of $P$ with eigenvalue
$1-\frac{2|S|}{d}$.  The probability that the random walk starting
form a random point in $B$ ends up in $B$ after $r$ steps equals
\begin{eqnarray*}
\Pr[Q_B\in B]&=&\frac{1}{|B|}\sum_{a,b\in B}
\left(P^r\right)_{ab}\\&=&\frac{2^d}{|B|}\langle P^r\1_B,\1_B\rangle\\
&=& \frac{2^d}{|B|}\sum_{S\subseteq\{1,\ldots,d\}}\widehat
{\1_B}(S)^2
\left(1-\frac{2|S|}{d}\right)^{r}\\
&\le& \frac{2^d}{|B|}\sum_{\substack{S\subseteq\{1,\ldots,d\}\\
|S|\le d/2}}\widehat {\1_A}(S)^2
\left(1-\frac{2|S|}{d}\right)^{r}\enspace,
\end{eqnarray*}
where we used the fact that $r$ is odd (i.e. we dropped negative
terms).

Thus, using~\eqref{eq:use} we see that
\begin{eqnarray*}
\Pr[Q_B\in B]\le
\frac{2^d}{|B|}\sum_{S\subseteq\{1,\ldots,d\}}\widehat
{\1_B}(S)^2\cdot e^{-2r|S|/c}\le \frac{2^d}{|B|}\cdot
\left(\frac{|B|}{2^d}\right)^{\frac{2}{1+e^{-2r/c}}}=\left(\frac{|B|}{2^d}\right)^{\frac{1-e^{-2r/c}}{1+e^{-2r/c}}}\enspace.
\end{eqnarray*}
\end{proof}

\begin{proof}[Proof of Proposition~\ref{prop:general}] Assume that $r$ is an odd
integer and  $R<\frac{d}{2}$. For $x\in \{0,1\}^d$ let $W_r(x)\in
\{0,1\}^d$ be the random point obtained by preforming a random walk
 for $r$ steps starting at $x$. Since $\|x-W_r(x)\|_1\le r$ we know
 that $\Pr\left[\h\left(W_r(x)\right)=\h(x)\right]\ge p$. Taking
 expectation with respect to the uniform probability measure on
 $\{0,1\}^d$ we deduce that
\begin{eqnarray}
p&\le& \E_{x\in
\{0,1\}^n}\Pr\left[\h\left(W_r(x)\right)=\h(x)\right]\nonumber\\
&=& \E_\h \Pr\left[x\in \{0,1\}^n:\ W_r(x)\in
\h^{-1}\left(\h(x)\right)\right]\nonumber\\
&=& \E_\h\sum_{k\in \mathbb N}\Pr\left[x\in \{0,1\}^n:\ W_r(x)\in
\h^{-1}\left(\h(x)\right)\ \wedge\
 \h(x)=k\right]\nonumber\\
 &=& \E_\h\sum_{k\in \mathbb N}
 \frac{\left|\h^{-1}(k)\right|}{2^d}\Pr\left[Q_{\h^{-1}(k)}\in
 \h^{-1}(k)\right]\nonumber\\
 &\le& \E_\h\sum_{k\in \mathbb N}
 \frac{\left|\h^{-1}(k)\right|}{2^d}\cdot\left(\frac{\left|\h^{-1}(k)\right|}{2^d}\right)^{\frac{e^{2r/d}-1}{e^{2r/d}+1}}\label{eq:useLemma}\\&=&
 \E_\h\E_{x\in
 \{0,1\}^d}\left(\frac{\left|\h^{-1}(\h(x))\right|}{2^d}\right)^{\frac{e^{2r/d}-1}{e^{2r/d}+1}}\nonumber\\
 &\le& \E_{x\in
 \{0,1\}^d}
 \left(\frac{\E_\h\left|\h^{-1}(\h(x))\right|}{2^d}\right)^{\frac{e^{2r/d}-1}{e^{2r/d}+1}}\label{eq:jensen}\\
 &\le&
 \left(q+e^{-\frac{1}{d}\left(\frac{d}{2}-R\right)^2}\right)^{\frac{e^{2r/d}-1}{e^{2r/d}+1}}\label{eq:useCoro}\enspace,
\end{eqnarray}
where in~\eqref{eq:useLemma} we used Lemma~\ref{lem:walk},
in~\eqref{eq:jensen} we used Jensen's inequality, and
in~\eqref{eq:useCoro} we used Corollary~\ref{coro:use}.
\end{proof}

\bigskip

\bigskip

\noindent{\bf Acknowledgements.} We are grateful to Jirka
Matou\v{s}ek for helpful suggestions.

\bibliography{lsh}
\bibliographystyle{abbrv}

\end{document}